\documentclass[a4paper,10pt]{article}

\usepackage[english]{babel}
\usepackage[utf8x]{inputenc}
\usepackage[T1]{fontenc}
\usepackage{booktabs}

\usepackage[a4paper,top=3cm,bottom=2cm,left=3cm,right=3cm,marginparwidth=1.75cm]{geometry}
\usepackage{amsmath}
\usepackage{graphicx}
\usepackage{placeins}
\usepackage{float}
\usepackage{setspace}
\usepackage[colorlinks=true, allcolors=blue]{hyperref}
\usepackage[numbers,sort&compress]{natbib}

\title{Dewetting Fingering Instability in Capillary Suspensions: Role of Particles and Liquid Bridges}
\author{Lingyue Liu$^{a, \dag, \ast}$, Mete Abbot$^{b, \dag}$, Philipp Brockmann$^{b}$, Ilia V. Roisman$^{b}$, \\
Jeanette Hussong$^{b}$, Erin Koos$^{a,\ast}$}
 \date{ \small
 {$^{a}$ KU Leuven, Department of Chemical Engineering, Celestijnenlaan 200J, Leuven, 3001, Belgium} \\
 $^{b}$ Technical University of Darmstadt, Institute for Fluid Mechanics and Aerodynamics, Peter-Gr\"{u}nberg-Str. 10, Darmstadt, 64287, Germany \\
$^{\ast}$ E-mail: lingyue.liu@kuleuven.be, erin.koos@kuleuven.be \\
$^{\dag}$ These authors contributed to this work equally \\}

\begin{document}

\maketitle

\begin{abstract}

This study investigates the fingering instability that forms during stretching of capillary suspensions with and without added nanoparticles. The dewetting process is observed using a transparent lifted Hele-Shaw cell. The liquid bridge is stretched under constant acceleration, and the resulting instability patterns are recorded using two high-speed cameras. Finger-like structures, characteristic of the Saffman-Taylor instability are observed. The total length of the dendrites and the intersecting number of branches are quantified. We reveal the roles of microparticles, nanoparticles, and the secondary liquid during the fingering instability.
The addition of microparticles to pure liquid enhanced finger length due to increased particle interactions and nucleation sites for bubbles. The addition of secondary fluid reduces fingering length by forming a strong interparticle network. Incorporation of nanoparticles induces an early onset of cavitation and enhances fingering instability. However, nanoparticles make the capillary suspensions' overall microstructure more homogeneous, reduce the sample variation in fingering patterns, and promote the even distribution of gel on both slides during splitting. 
These findings highlight the complex interactions governing dewetting in capillary (nano)suspensions. This knowledge has potential applications in microfluidics, 3D printing, and thin-film coatings, where controlling dewetting is crucial. \\


\end{abstract}





 

\section{Introduction}

Capillary suspensions, consisting of a bulk liquid phase, solid particles, and a secondary immiscible liquid that forms interparticle capillary bridges, exhibit a unique viscoelastic behavior with significant applications for advanced materials~\cite{weis2020,nider2024}. These complex structures enable the development of low-fat foods, porous ceramics with controlled properties, and printing inks~\cite{koos2012,dittmann2013,hoffmann2014}.
In these applications, material properties are governed by complex flow phenomena arising from interactions between solid particles, liquid phases, and capillary forces. The dewetting dynamics under stretching conditions play a crucial role in droplet formation, film stability, and sample deposition in printing or coating operations. Both bulk and secondary liquids, combined with micro- and nanoparticles, lead to an intricate interplay between capillary forces, viscous forces, and particle interactions~\cite{Haessig2025,brulin2021}.

A liquid bridge is a thread of fluid that forms between two solid surfaces, controlled by surface tension forces that cause the liquid to adopt a curved shape, spanning the gap~\cite{orr1975,plateau1857}. The stability of such a bridge is influenced by several factors, including the volume of liquid, the characteristics of the surfaces, and the surrounding environmental conditions. The fundamental behavior of liquid bridge stretching has been extensively studied for Newtonian fluids~\cite{plateau1857,orr1975,mckinley2002,anna2001,brulin2020,brulin2020b,asghar2023,lopez2023,moffatt2021}. The basic case of liquid bridge stretching consists of a Newtonian liquid and two parallel plates, with one plate moving. The liquid bridge stretching is studied extensively using the Capillary Breakup Extensional Rheometer (CaBER), with a dimensionless height $1 < \lambda = H_0/D_0 <2 $, where $H_0$ is the initial height and $D_0$ is the initial diameter of the liquid bridge. The capillary thinning and breakup dynamics depend on the viscosity, surface tension, and elastic properties of the fluid~\cite{anna2001,mckinley2002}. When the liquid bridge is confined in a thin gap with dimensionless height $\lambda \ll 1$, this experiment setup resembles a lifted Hele-Shaw cell with the formation and breakup of a ligament between two liquid menisci. The commonly studied lifted Hele-Shaw experiments consider Newtonian fluids and constant lifting velocities of a few microns per second, where the inertial effects are small~\cite{amar2005,dias2013}. During stretching, the radius of an axisymmetric bridge evolves to maintain volume conservation and can be described as $R^2(t)H(t)=R^2_0H_0$ for perfect cylindrical bridges. During stretching, the circular interface can become unstable due to the high interface retraction rate and negligible height~\cite{brulin2020,brulin2021}. The perturbed interface can then be described as $\mathcal{R}(\theta, t)=R(t)+\zeta(\theta, t)$, where $\theta$ is the azimuthal angle, $R(t)$ is the time-dependent radius of the unperturbed menisci, and the $\zeta(\theta, t)$ is the net interface perturbation.

In addition to interfacial instabilities, the rapid stretching of confined liquid bridges can trigger cavitation, which has distinct characteristics in a Hele-Shaw cell geometry compared to bulk systems. Due to the effects of confinement, bubbles preferentially expand in the planar direction rather than spherically. With decreasing gap height, the bubble lifetime increases by up to 50~\% while the maximum projected bubble radius remains constant~\cite{Quinto-su2009}. At high separation velocities, cavitation appears and the force response shows a plateau during which bubbles grow. When the bubbles eventually break, outside air rapidly enters the cavity, causing an abrupt force drop~\cite{Poivet2003}. The presence of microscopic nuclei can lead to heterogeneous nucleation at pressures exceeding the theoretical cavitation threshold~\cite{Chen2020}. This interplay between cavitation and fingering becomes particularly complex in Hele-Shew cells, as both phenomena compete for stress relaxation during stretching via fingering through edge effects and cavitation through bulk deformation. 

While the fundamental behavior of liquid bridges has been well-characterized for simple systems, real-world applications often involve additional complexities such as additional particles. In capillary suspensions,  microparticles form a sample-spanning network by aqueous bridge connections at the micrometer scale, with oil as the bulk phase rather than air. Understanding these particle-liquid bridge interactions is crucial for applications ranging from materials processing to printing technologies and the dynamics of particle-liquid bridge interactions have been extensively studied at the single-particle scale using millimeter-sized particles. When a particle impacts a wet surface, it first penetrates the liquid layer, generating elastic waves in both the particle and wall. These waves, along with potential plastic deformation at higher impact velocities, influence the initial particle rebound. During rebound, a liquid bridge forms, and the stresses in this stretching bridge determine the final particle velocity and potential substrate deposition~\cite{Cruger2016}.
The ratio between rebound and impact velocities, defined as the restitution coefficient, characterizes the energy dissipation during these collisions~\cite{Muller2016}. Grohn \emph{et al.}~\cite{grohn2022} measured the restitution coefficient and bridge breakage for varying liquid volumes, finding that more liquid leads to longer rupture duration. Additional studies~\cite{ma2016,buck2018} showed that in oblique impacts, wet particles exhibit higher tangential restitution coefficients than dry ones due to lubrication effects. The model presented by Davis~\cite{davis2019} extended this to multi-wet body collisions, predicting outcomes from bouncing to clumping based on the interplay of inertia, viscosity, and surface tension forces.

The complexity of liquid bridge behavior increases substantially when both micro and nanoparticles are present in the system. Based on our previous studies, we discovered that capillary nanosuspensions are normal capillary suspensions with a low volume fraction of nanoparticles inside the secondary liquid. In ceramic systems, these nanoparticles can strengthen sintering necks and form dense bridges~\cite{weis2020,dittmann2013}. However, in wet precursor systems, nanoparticle addition decreases the yield stress, enhances interparticle movement, and promotes secondary liquid distribution, making the network more compressible and homogeneous~\cite{weis2020,dittmann2013,liu2024,Aksoy2023}.

This study investigates the dynamic behavior of capillary nanosuspensions under the combined influence of thin-gap stretching and high acceleration. We employ a novel experimental setup, incorporating high-speed imaging, to directly observe the collective dewetting process. Our findings provide crucial insights into the dewetting dynamics of capillary nanosuspensions, with potential implications for controlling their rheological properties and optimizing their use in applications such as 3D printing and thin-film fabrication. The dynamic stretching of capillary nanosuspensions can lead to complex dewetting phenomena, including the formation of fingering instabilities. These instabilities arise from the interplay of capillary forces, viscous forces, and the presence of particles, making their accurate prediction a significant challenge. During fast stretching, the mass balance is not easily determinable even for Newtonian fluids, as the plates remain wetted by a thin residual liquid layer when the meniscus recedes~\cite{brulin2021}. The complexity of the system and high acceleration bring challenges in modeling the fingering instability mathematically, meanwhile, this work focuses on the fingering number, total dendritic length changes, and cavitation effect caused by additional interparticle bridges and nanoparticles. The findings show the complex interplay between the properties of the microparticles, nanoparticles, liquid bridges, and instabilities in the flow within the confined gap.


\section{Materials and methods}

\subsection{Sample preparation}

The capillary suspensions were prepared using a bulk phase consisting of a blend of 83.8 vol\% 1,2-cyclohexane dicarboxylic acid diisononyl ester (Hexamoll DINCH, BASF) and 16.2 vol\% \textit{n}-dodecane ($>$ 99\%, Sigma-Aldrich). This mixture possesses a density of 0.91~g/ml, a viscosity of 21~mPa$\cdot$s, and a surface tension of 28.4 $\pm$ 0.2~mN/m (Attension, Biolin Scientific). With equilibrated full coverage of 10~$\mathrm{\mu}$m silica microparticles at the bulk liquid interface, this surface tension of the oil mixture decreases from 28.4~mN/m to 22~mN/m. The secondary liquid phase comprised a mixture of 50 vol\% glycerol ($>$ 99.5\%, Carl Roth) and 50 vol\% ultrapure water (Arium 611DI, Sartorius Stedim Biotech) with a density of 1.12~g/ml and a viscosity of 7~mPa$\cdot$s. The aqueous glycerol has a surface tension of 68.3 $\pm$ 0.3~mN/m, and an interfacial tension with bulk liquid of 25.5 $\pm$ 0.7~mN/m. The silica microparticles (SOLAD nonporous PNPP10.0NAR Glantreo), with an average diameter of 10 $\mathrm{\mu}$m and a polydispersity of 5~\%, fluorescently labeled using rhodamine B isothiocyanate (RBITC, Sigma-Aldrich). The three-phase contact angle of the hydrophilic microparticles at the liquid-liquid interface is 50$^{\circ}$, as determined in our previous work~\cite{liu2024} using confocal images of the particles at the liquid-liquid interface in a microchannel~\cite{allard2022}. The nanoparticles are produced via St\"{o}ber synthesis, with an average size of 150 $~\pm$ 12~nm. Nanoparticle-loaded samples were prepared at a volume fraction of $\phi_{NP}$ = 1 vol\%, relative to the secondary liquid fraction $\phi_{sec}$. The addition of this amount of nanoparticles does not influence the interfacial tension nor the viscosity of the base (secondary) liquid. The bulk liquid is chosen to match the refractive index of the silica microparticles ($n$ = 1.455)~\cite{allard2022} for better optical observation, besides, their low viscosity and moderate interfacial tension allow the formation of stable liquid bridges. The refractive index of the secondary liquid is slightly mismatched (1.4). The secondary liquid is chosen to extend shelf life of using pure water, and high volume fraction of glycerol is too viscous to create uniform bridging. Silica microparticles were used
due to their well-defined size, shape, and smoothness, enabling controlled particle interactions. Furthermore, their fluorescent labeling facilitated visualization of the dewetting process.

To achieve capillary suspensions with a secondary liquid volume fraction of $\phi_{sec}$ = 1 vol\%, 10.1 $\mathrm{\mu}$l of the secondary liquid (either pure aqueous glycerol or the nanosuspension), was added to 800 $\mathrm{\mu}$l of the bulk liquid. The liquids underwent a two-step mixing process using a 3.175~mm diameter ultrasonic horn (Digital Sonifier model 450, Branson Ultrasonics Corporation). Initially, the fluids were dispersed at 10\% amplitude for 10~s followed by 30\% amplitude for 30 s. Subsequently, 435 mg of silica microparticles ($\phi_{MP}$ = 20 vol\%) were introduced into the sample. After particle addition, the sample underwent additional mixing using the ultrasound horn at 10\% amplitude for 20~s.

The suspension is simply prepared by manually mixing the same amount of silica microparticles with bulk liquid using a spatula without the addition of the secondary liquid.

\subsection{Experimental setup}
 
\begin{figure}[tbp]
    \centering
      \includegraphics[width=0.7\textwidth]{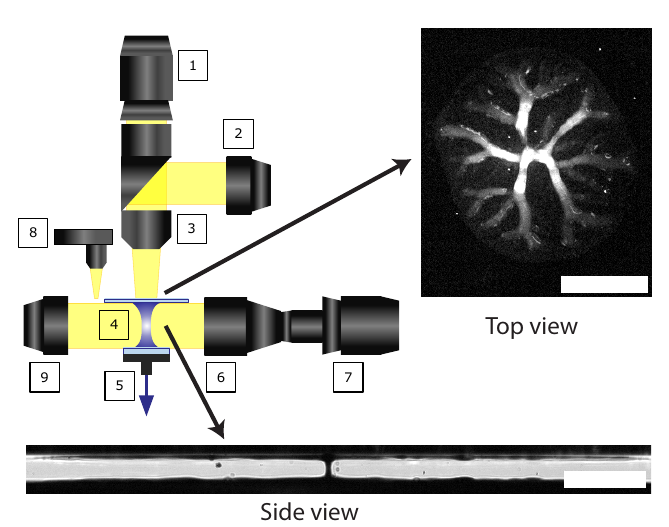}
      \caption{Schematic configuration of the liquid bridge stretching experiment (1) High-speed camera, (2) High Power LED, (3) Navitar Zoom Objective, (4) Gel/Liquid placed between an upper and lower substrate, (5) Linear actuator, (6) Telecentric objective, (7) High-speed camera (8) Confocal sensor (9) Telecentric lighting. The scale bars on the exemplary images are 5~mm.}
      \label{fig:setup}
    \end{figure}

Experiments performed within this study are based on a rapidly stretched liquid bridge forming between two horizontally aligned substrates. The stretching process is induced by the motion of the lower substrate with the upper substrate remaining fixed. A sketch of the experimental setup is shown in Fig.~\ref{fig:setup}. 
The main component of the stretching apparatus is the linear drive (Akribis, type SGL100-AUM3) operated by a servo controller (Metronix, ARS2108), providing reliable accelerations up to 180 m/s$^2$. The linear unit has a peak force of 900 N and a positioning accuracy of 5~$\mathrm{\mu}$m.

Two separate high-speed video systems are utilized for observation. The first high-speed camera (Phantom T1340, labeled ``1'') with a pixel size of 27 $\times$ 27~$\mathrm{\mu}$m records the liquid bridge from the top through the transparent upper substrate, where the bright pixels represent the sample. A frame rate of 1000 fps and a resolution of 1024 px $\times$ 976 px is chosen for the experiments with a spatial resolution of 50~$\mathrm{\mu}$m/px. Illumination and magnification are achieved using a custom-built microscope comprising an optical tube (InfiniTube Special, Infinity Photo-Optical) illuminated by a 7 W high-power green LED ($ \lambda \approx 532$ nm, ILA iLA.LPS v3, labeled as ``2''). A dichroic mirror and two bandpass filters, mounted on a filter cube (Thorlabs DFM1/M, labeled as ``9''), direct the excitation light through the objective and filter the fluorescence signal emitted by the excited particles. The system employs an Infinity Photo-Optical IF-3 objective (``3'') with 1$\times$ magnification. The images generated with a high-speed camera (``1'') are referred to as the top view in the following. The second high-speed camera (Phantom Miro Lab 110, labeled ``7'') with a pixel size of 20 $\times$ 20 $\mathrm{\mu}$m records the side profile using a telecentric objective (``6'') and a telecentric lighting (``9''), the images generated with this camera are hereafter referred to as side view, where the bright pixels are the background and the dark pixels are the substrates and samples in between. The camera is operated with a frame rate of 3200 fps and a resolution of 800 $\times$ 640 px with a spatial resolution of 50 $\mathrm{\mu}$m/px. The distance between the substrates is measured with a confocal displacement sensor (Micro Epsilon IFS2407-0.1, ``8'') with an accuracy of up to 3~nm within a maximum measuring range of 0.1~mm. The gap of the experiments in this study is set to 50~$\mathrm{\mu}$m.

To ensure precise parallel alignment of the two substrates (``4''), an automated procedure was implemented. A confocal displacement sensor was used to measure the distance between the substrates at three different locations. From these measurements, the angle between the plates in both the x- and y-planes was calculated. If any misalignment exceeding 0.01$^\circ$ (50~nm) was detected, piezo motors (Thorlabs PIA50) adjusted the substrate orientation accordingly. This process was repeated iteratively until the desired parallelism was achieved. Once the plates were sufficiently parallel, the liquid droplet was deposited between them.

The constancy of sample loading is of great importance in our experiments. For normal suspensions with no liquid bridges and pure particles, we manually stirred the sample with a spatula before 50~$\mathrm{\mu}$l volume was taken by the syringe and dosed onto the bottom plate. For capillary (nano)suspensions, the exact dosing is impossible, since the compression forces can expel the bulk liquid and increase the volume fraction. The gels were manually scooped out and spread onto the bottom plate, then the bottom plate was slowly brought to the desired position/gap without compressing the samples extensively.


\subsection{Image processing}

\begin{figure}[hbt]
    \centering
      \includegraphics[width=0.7\textwidth]{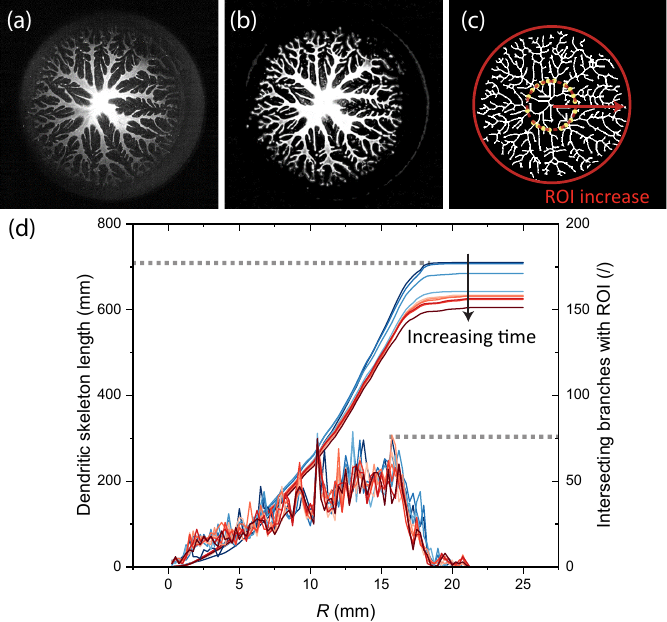}
      \caption{(a-c) Exemplary images of the top view dendritic classification method. (d) Dendritic skeleton length and intersecting branches with the region of interest (ROI) over $R$, the increasing radius of ROI, the intersection of branches with ROI are shown in yellow.}
      \label{fig:imageprocessing}
    \end{figure}

The fingering patterns were extracted from the background using the machine learning-based image processing software Ilastik. This software employs a combination of edge detection, pixel intensity, and texture features to classify pixels based on provided training, effectively separating sample traces from the glass plates (Fig.~\ref{fig:imageprocessing}a). 
After training Ilastik to differentiate fingers from the background, the processed images (Fig.~\ref{fig:imageprocessing}b) were then binarized and skeletonized using a self-written MATLAB code. Starting from the center of mass of the first processed image (Fig.~\ref{fig:imageprocessing}c), a region of interest (ROI) was defined and expanded with each increment in a ring radius ($R$) of 5 pixels (250~$\mathrm{\mu}$m). This 5-pixel increment was determined experimentally to reduce noise while preserving valuable information. The pixels within the skeletons were tallied as the ring radius increased, serving as an indicator of the total area of the branches, named as dendritic skeleton length (50~$\mathrm{\mu}$m/pixel). Analysis and plotting of the images extended from the first to the tenth frame, the total dendritic skeleton lengths are plotted. Besides, the numbers of intersecting branches with ROI are also shown as the numbers of skeletons intersecting with the ROI ring peripheries, as shown by the yellow dots in Fig.~\ref{fig:imageprocessing}c.

Due to the diverse pattern formation mechanisms observed, the frame selection criteria were adapted to the dominant behavior in each sample. For cavitation-dominant samples, where multiple disconnected void structures emerge, we analyzed the frame where cavitation had stabilized and distinct voids were visible. In contrast, for instability-dominant samples, which typically show connected dendritic patterns converging toward the center, we selected the frame exhibiting the maximum total dendritic length. This skeletonization method transforms branches of varying thicknesses into unity-pixel skeletons, providing a standardized measure of pattern length regardless of branch thickness. This approach accounts for both connected and isolated segments, as the concept of a main skeleton is primarily applicable to instability-dominated samples and not meaningful for cavitation-dominated cases where multiple independent structures form.

\section{Results and discussions}

\subsection{Suspension dewetting}\label{sec: suswetting}

\begin{figure}[tbp]
    \centering
      \includegraphics[width=0.5\textwidth]{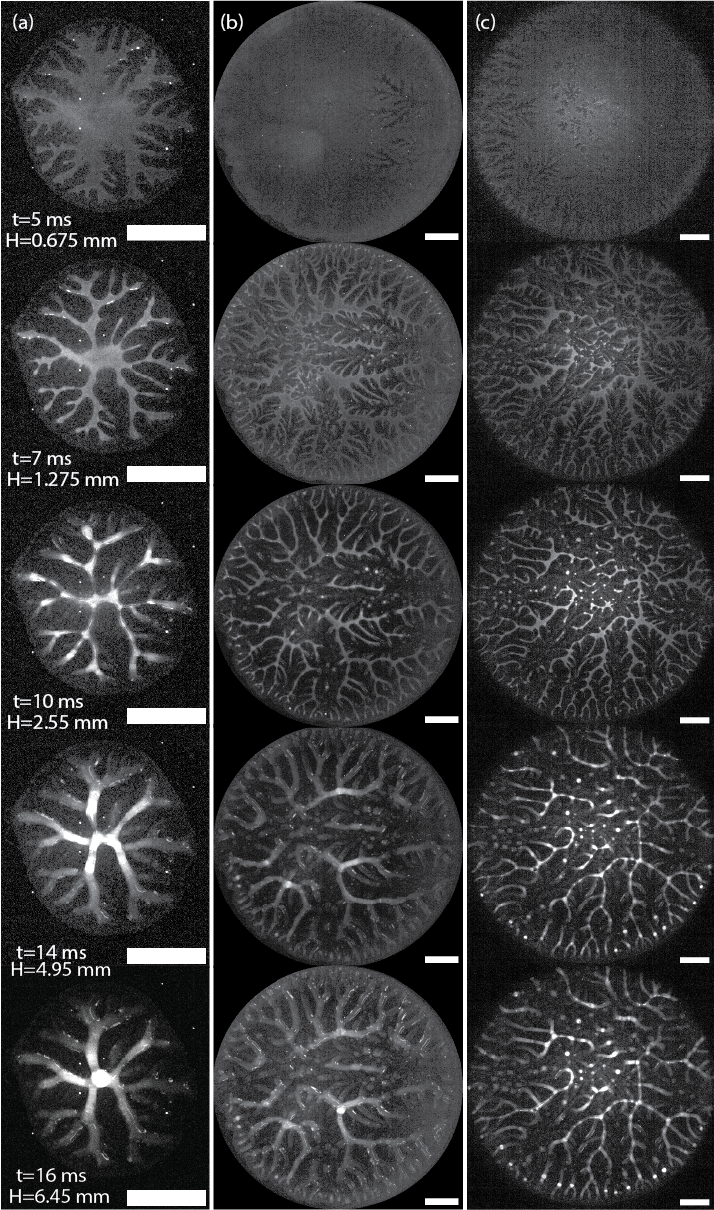}
      \caption{Three suspensions of (a) small initial diameter 12~mm, and larger initial diameter 36~mm where (b) fingering instability or (c) cavitation is dominant, at different timestamps, dewetted with an acceleration of 50~m/s$^{2}$. The gap height at each timestamp is $H$. The scale bars are 5~mm.}
      \label{fig:noBridges_top}
    \end{figure}

As shown in Fig~\ref{fig:noBridges_top}a, 
the fingering pattern for a low sample volume (12~mm initial diameter) resembles that of a Newtonian fluid under high substrate acceleration, as observed by Brulin \emph{et al.}~\cite{brulin2020}. Dewetting begins at $t = 5$~ms, the liquid begins to dewet the top plate, and the lost volume can only be provided by the outer due to volume conservation. Intensified fluorescent signals indicate particle accumulation at the fingering edges. At $t=7$~ms, the dendritic branches elongate vertically, causing fluid to accumulate along the vertical direction of the branches. This accumulation increases the concentration of particles in the vertical region, leading to a more pronounced fluorescence signal.

During dewetting, for particles to be suspended and follow the flow, the lift force $F_L$ must exceed the gravitational force $F_G$. The gravitational force $F_G = \frac{1}{6} \pi d_p^3  \Delta \rho $, where $d_p$ is the particle diameter and $\Delta \rho$ is the density difference. The lift force for particles near the plate can be expressed as seen in Eq.~\ref{eq:simplified_lift}~\cite{li2022},
    \begin{equation}
            F_L=\left(\frac{3}{2}\right)^{\frac{3}{2}} K\left(1-\frac{d_p}{2 H_0}\right) \sqrt{1-\frac{d_p}{H_0}} {d_p}^3 \sqrt{\frac{\rho_f \eta_f U^3}{H_0^3}}
            \label{eq:simplified_lift}
    \end{equation} 
where $K = 81.2$ is the Saffman force coefficient, $\rho_f$ is the fluid density, $\eta_f$ is the dynamic viscosity of the base fluid, and $U$ is the average velocity of liquid interfaces. When taken $d_p$ = 10$~\mathrm{\mu}$m and $H_0$ = 50$~\mathrm{\mu}$m, the critical retraction speed is $U =$ 90$~\mathrm{\mu}$m/s. In our experiments, the observed retraction speeds (including the meniscus radius change rate, flow along branches, and branch thinning rate) all exceed this value. As shown in Fig~\ref{fig:noBridges_top}a, no particles are found deposited on the surface where the air invaded (several fixed dots observed within all the images are contaminants on the objective lens). 

Affirming that the flow velocity exceeding the critical velocity particles require to suspend in the liquid, the intensity of fingering patterns can be used to identify the microparticle displacement. At $t = 7$~ms, branches are brighter at the edges. The change in the position of the discontinued bright spots indicates the movement of  fluorescently labeled microparticles from edges towards the knots at $t = 10$~ms. At $t = 14$ - 16~ms, the intensified signal toward the middle region indicates the directional particles migration, forming a particle-laden major filament.

For higher sample volumes with an initial diameter of 36~mm (Fig.~\ref{fig:noBridges_top}b and c), the dewetting patterns differ significantly. The local stress differences caused by fast flow can either be compensated by instability and radius decreasing or cavitation bubble appearance and growth~\cite{Poivet2003}. In both Fig.~\ref{fig:noBridges_top}b and~\ref{fig:noBridges_top}c, cavitation begins within the initial 5~ms of stretching, typically driven by the presence of foreign particles decreasing the total free energy needed for bubble nucleation~\cite{Chen2020,deng2022,blander1975}. It was observed that the cavitation formation appears almost simultaneously, within a submicrosecond. In Fig.~\ref{fig:noBridges_top}b, the Saffman–Taylor instability dominates with the minimum effect of cavitation, and the branches are mostly interconnected ($t = 10$~ms). For the sample shown in Fig.~\ref{fig:noBridges_top}c ($t = 5$~ms), the cavitation begins with the appearance of instability. As cavitation and instability interface collide, the connected networks break. One interesting observed phenomenon is that the branches start to widen between $t = 10$ - 16~ms, this is especially visible for Fig.~\ref{fig:noBridges_top}a and ~\ref{fig:noBridges_top}b. At the initial stage, the thickness of fingering instability branches is primarily driven by the extreme confinement effect of the small gap height ($\lambda = H_0/D_0 <1$). Due to the small initial heights between plates, the liquids flow in a quasi-2D plane and the shear near the wall is maximum, this results in flows dominated by viscous forces and high pressure gradient caused thin fingers (with vertical filaments) connecting both plates~\cite{Poivet2003}. After the filaments break in the middle due to the stretching, surface tension causes the remaining liquid to retract and redistribute on each plate, leading to an apparent widening of the branches as the fluid relaxes and spreads laterally.

\begin{figure}[tbp]
    \centering
      \includegraphics[width=0.6\textwidth]{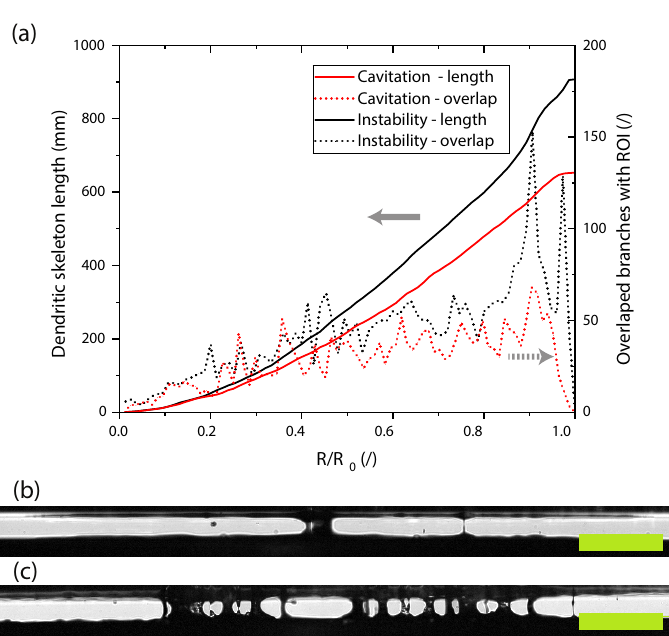}
      \caption{Total dendritic skeleton length and intersecting branches with ROI of suspensions of larger initial diameters 36~mm where dewetted with an acceleration of 50~m/s$^{2}$. Their side views are as shown, where in (b) fingering instability dominates and in (c) cavitation dominates. The scale bars are 5~mm.}
      \label{fig:noBridge_cavitation_area_overlap}
    \end{figure}

The experimental values for the total dendritic skeleton length and the intersecting branch number (skeletonized branch number values on ROI) between instability-dominated (Fig.~\ref{fig:noBridges_top}b) and cavitation-dominated (Fig.~\ref{fig:noBridges_top}c) samples are shown in Fig.~\ref{fig:noBridge_cavitation_area_overlap}a. 
The instability-dominated case shows a nearly 40~\% increase in branch length and a 100~\% increase in intersecting branch number. This phenomenon agrees with the estimation of Brulin~et~al.~\cite{brulin2021}, where the cavitation lowers the branch number significantly (up to 120\%). The broken branches caused by cavitation and instability eventually retract, leading to small particle-laden bridges that appear as single dots in Fig.~\ref{fig:noBridges_top}c. Their side view differences shown in Fig~\ref{fig:noBridge_cavitation_area_overlap}b and \ref{fig:noBridge_cavitation_area_overlap}c, where instability dominant pattern retracts inward and forms limited amount of thick filaments. Meanwhile the cavitation-dominanted pattern spreads outward due to the extensive amount of freshly nucleated bubbles to form multiple single filaments during stretching. The transition between domination types was unable to be clearly predicted but likely to be associated with microparticle assembly due to compression, inducing the onset of cavitation.

\subsection{Capillary suspension dewetting}

\begin{figure}[tbp]
    \centering
      \includegraphics[width=0.6\textwidth]{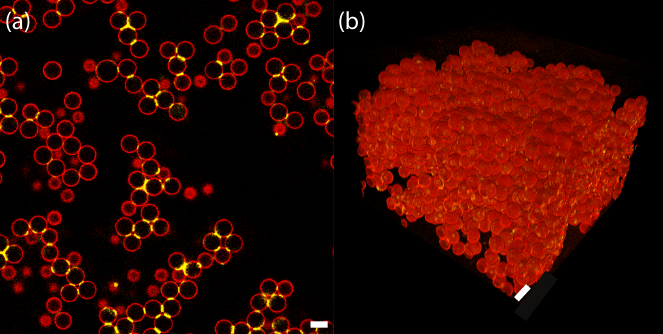}
      \caption{Confocal micrographs of capillary suspensions without nanoparticles in (a) 2D and (b) 3D, the red spheres are microparticles and yellow parts represent liquid bridges. The scale bars are 10~$\mathrm{\mu}$m.}
      \label{fig:CapS_confocal}
    \end{figure}

In the capillary suspension system, particles are internally connected by the aqueous liquid bridges (Fig.~\ref{fig:CapS_confocal}a), 
creating a sample-spanning network (Fig.~\ref{fig:CapS_confocal}b). The gravitational force $F_G$ becomes less significant as the capillary force $F_{cap}$ provided is generally several orders of magnitude larger~\cite{bossler2016}. The total lifting force applied to the Newtonian bridge cross-section at the middle plane is affected by an interplay between viscous force, inertia stress, and capillary force~\cite{roisman2023}. The microstructure of this gel system has been studied extensively~\cite{liu2024,allard2022,bossler2018, bossler2017,koos2011}, when the secondary liquid volume becomes excessive, the microparticles aggregate and cluster extensively, forming an inhomogeneous particle structure. Eventually, the behavior of the clusters (fully wetted microparticles) can be analogically compared to the suspension stretching explained previously in Sec~\ref{sec: suswetting}. During stretching, the sparsely connected particles can collide with the clusters, which eventually increases the energy dissipation and decreases the coefficient of restitution, leading to a collective unidirectional displacement~\cite{ma2016,buck2018}.

\begin{figure}[tbp]
    \centering
      \includegraphics[width=0.3\textwidth]{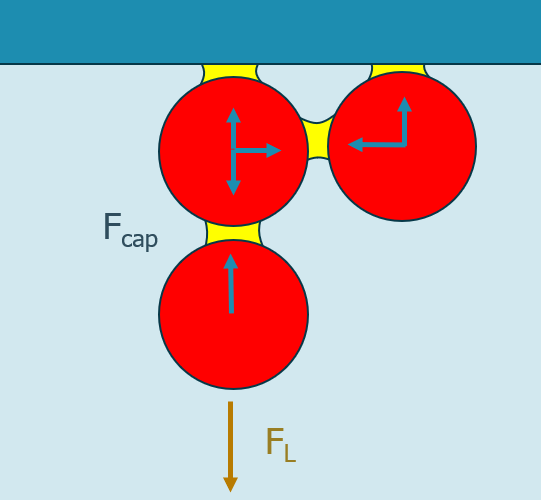}
      \caption{Illustration of capillary suspensions being stretched where $F_{cap}$ is the capillary force according to eq.~\ref{eq:capForce} and $F_L$ is the lift force according to Eq.~\ref{eq:simplified_lift}}
      \label{fig:ForceBalanceCapS}
    \end{figure}

During fast stretching, the interplay between viscous stress and inertia can be reflected by the Reynolds number. The Reynolds number between two accelerating microparticles is defined as Eq.~\ref{eq:bridgeReynolds}~\cite{brulin2020b}, 
    \begin{equation} \label{eq:bridgeReynolds}
        \mathrm{Re} \equiv \frac{\rho a^{1 / 2} k^{1 / 2} R_{bridge}^{3 / 2}}{\eta}
    \end{equation}
where $a$ is the acceleration, $R_{bridge}$ is the bridge neck radius, $k$ is the dimensionless ratio between the initial height and diameter of the radius. Based on our previous work~\cite{liu2024}, the single bridge size and the maximum rupture distance between two microparticles of 10~$\mathrm{\mu}$m are 2-5~$\mathrm{\mu}$m and 2~$\mathrm{\mu}$m respectively, the maximum dimensionless height $k_{max}$ is 1. The resulting Reynolds number is roughly 0.01 $\ll$ 1. At small Reynolds numbers, viscous stress plays an important role during the entire stretching and the inertia effect becomes insignificant~\cite{brulin2020b}. The viscous contribution of the force can be expressed as $F_{\eta}=3\pi \eta R_{bridge}^2 u_z$, where the maximum particle velocity in $z$ axis $u_{z,max} = a t$, $a$ is the substrate acceleration, and $u_{z,max}$ does not exceed the substrate velocity. Since the viscosity of the bulk liquid is larger than that of the secondary liquid, the viscous contribution comes from the bulk liquid ($\approx$10$^{-11}$~N)~\cite{Haessig2025,lopez2023}, which is negligible in comparison to the capillary force ($\approx$10$^{-6}$ N), as we reported previously~\cite{liu2024}. As shown in Fig~\ref{fig:ForceBalanceCapS}, the lift force $F_L$ can either be as described in Eq.~\ref{eq:simplified_lift} or $F_{cap}$, depending on whether the connecting particle is present.

\begin{figure}[tbp!]
    \centering
      \includegraphics[width=0.45\textwidth]{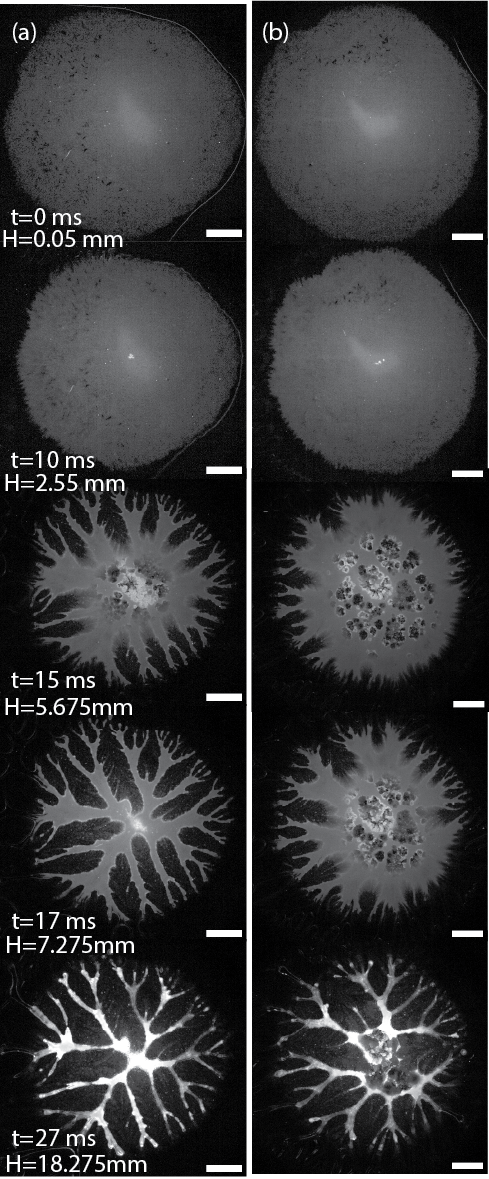}
      \caption{Capillary suspensions without nanoparticles at different timestamps without (a) and with (b) cavitation in their final pattern, dewetted with an acceleration of 50~m/s$^{2}$. The gap height at each timestamp is $H$. The scale bars are 5~mm.}
      \label{fig:1Bridges_top}
    \end{figure}

One interesting phenomenon observed in the stretching of capillary suspension is that the general radius $R_{0}$ starts to decrease uniformly as it is stretched, the sample area shrinks inwards of up to 13\% before cavitation occurs ($t = 10$~ms), as can be visually seen in Fig.~\ref{fig:1Bridges_top}. 
The normalized area $A$ (by initial area $A_0$) and the shrinkage rate $\partial A/ \partial t$ are shown in Supplementary Fig.~S1a and S1b, with an average shrinkage of 9~\% and a rate of 2.5~\%/ms respectively. When uniform shrinkage disappears the pressure drop cannot be compensated by the insufficient sample flow, filling up inwards from the outer rim, that is when cavitation begins~\cite{brulin2020,asghar2023,Poivet2003,blander1975}. Meanwhile, capillary suspension microstructure is strongly loading-dependent, and local microparticle aggregation can easily induce pre-onset cavitation, similar to the suspension situation shown in Sec~\ref{sec: suswetting}.

The capillary suspension fingering instability (Fig.~\ref{fig:1Bridges_top}) resembles the suspension pattern in Fig~\ref{fig:noBridges_top}. In suspension samples, interparticle forces are limited, with microparticle movement dominated by the lifting force $F_L$. Once the liquid-air interfaces are created by cavitation and by the Saffman–Taylor instability from the outer encounter, numerous fine filaments are created (Fig~\ref{fig:noBridge_cavitation_area_overlap}c), without the attractive forces leading to emergence. For suspensions, the instability and cavitation last roughly 5~ms (Fig~\ref{fig:noBridges_top}, $t = 5$ to 10~ms), while the same process in capillary suspensions takes longer due to the influence of capillary forces (Fig.~\ref{fig:1Bridges_top}b, $t = 10$ - 17~ms).

When experimenting with a Newtonian fluid, the bubbles that appear in the middle can either expand quickly, leading to cavitation structures that resemble a Voronoi diagram, or they fall below the critical diameter and disappear after the stress relaxation~\cite{brulin2021,brulin2020}. In capillary suspension case, the interparticle capillary force can be expressed as 
    \begin{equation} \label{eq:capForce}
        F_{cap}=\pi d_p \Gamma \cos \theta
    \end{equation}
where $\Gamma$ is the interfacial tension between the liquid phase and $\theta$ is the contact angle, which are 25.5 $\pm$ 0.7~mN/m and 50{\textdegree}, respectively~\cite{allard2022,liu2024,bindgen2022}. The resulting toroidal bridge capillary force is about 500~nN between two spherical particles of 10~$\mathrm{\mu}$m. As fingering instability proceeds, the attractive network eventually converges in the middle and covers up the cavitation pattern (Fig.~\ref{fig:1Bridges_top}a, $t = 15$ - 17~ms), resulting in one single large bridge in the center, indicating an instability controlled pattern. However, samples of similar sizes can also be cavitation-dominanted due to the loading variation caused heterogeneity inside the sample. As can be seen in Fig.~\ref{fig:1Bridges_top}b, $t = 15$~ms, the voids grow at a much faster rate at the same timestamp in Fig.~\ref{fig:1Bridges_top}a, and they cannot be readily compensated, resulting in a more complicated cavitation-dominated residual pattern at the center region.

Capillary suspensions show significant differences in dendritic skeleton length and intersecting branch number compared to pure suspensions (Supplementary Fig.~S2). In regular suspensions, cavitation separates fingering patterns into smaller segments, which further contract due to liquid retraction. However, in capillary suspensions, the strong interparticle capillary forces maintain network stability, opposing displacement of the newly created surfaces. This internal yield stress results in an elevated fingering instability caused by cavitation. The total fingering length of the sample with no clear cavitation decreases by nearly 43~\%, and its peak intersecting number decreases by 66~\%. As seen from Fig.~\ref{fig:noBridges_top}c, cavitation meets instability and the liquid at created discontinuous interfaces retracts mostly into single dots. However, due to the strong network, the gel structures at the interfaces do not shrink, inducing a complicated pattern near the center region and increasing the skeleton length (Fig.~\ref{fig:1Bridges_top}b).

\subsection{Capillary nanosuspension dewetting}

\begin{figure}[tbp]
    \centering
      \includegraphics[width=0.4\textwidth]{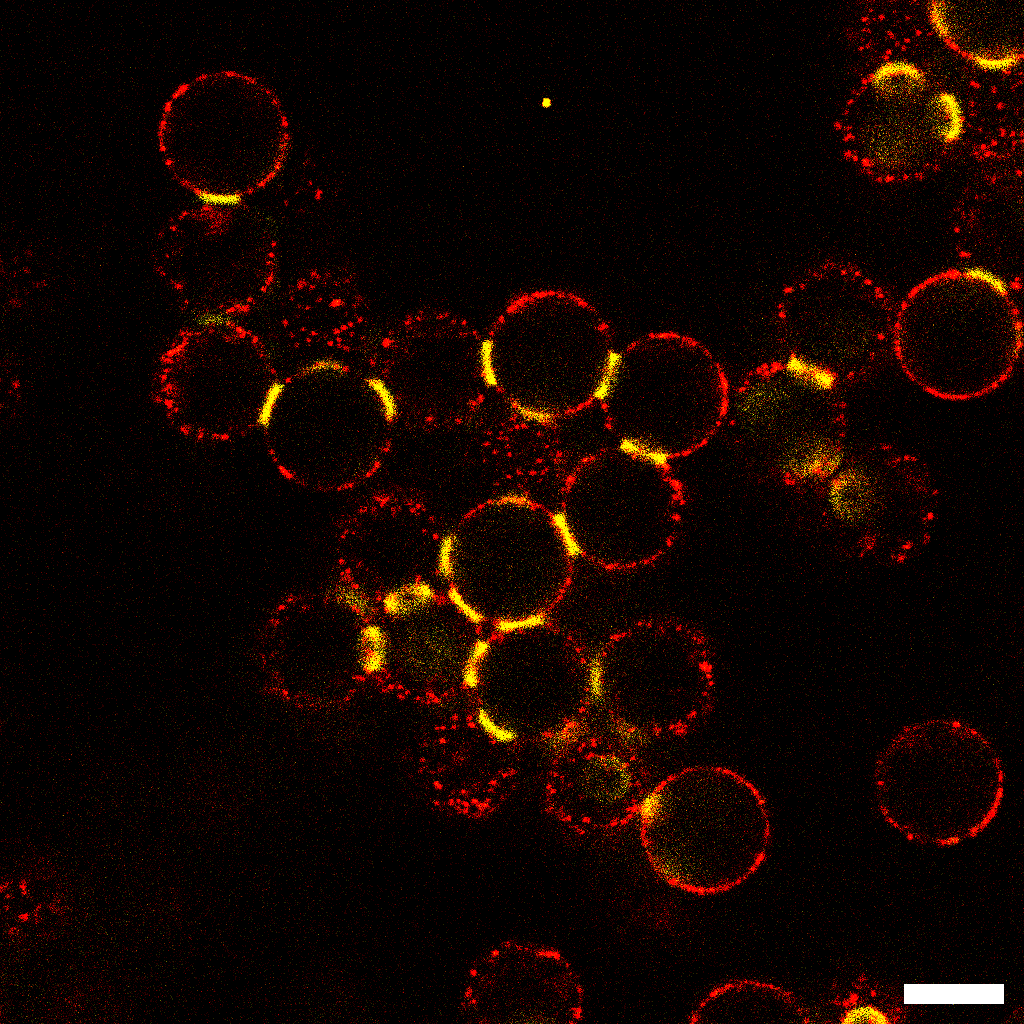}
      \caption{Confocal micrographs of capillary nanosuspensions ($\phi_{MP}$ = 20~vol\%, $\phi_{sec}$ = 1~vol\%, $\phi_{NP}=0.01$~vol\%). The nanoparticles, added in the secondary liquid phase are core-shelled fluorescent and hydrophilic, shown in red, the liquid bridges are shown in yellow, and the microparticles are undyed. The scale bar is 10~$\mathrm{\mu}$m.}
      \label{fig:NP_CapS}
    \end{figure}

In our previous study where we compressed the capillary (nano)suspensions~\cite{liu2024}, we observed that incorporating even a small amount (0.01~vol\%) of hydrophilic nanoparticles resulted in their uniform distribution on the surface of the microparticles. As shown in Fig~\ref{fig:NP_CapS}, 
these nanoparticles are proven to induce the formation of nanometer-sized liquid films~\cite{Fusco2024}. These films increase the restitution coefficient, slowing down the energy absorption upon particle collision, lubricating the relative movement, and increasing bridge instability during stretching~\cite{grohn2022,ma2016,buck2018}. As a result, the elastic component, storage modulus $G'$ decreases by a half order of magnitude due to the elimination of Hertzian contact force, this decreases the contact force and connectivity between microparticles during stretching. The overshoot for the viscous component, the loss modulus $G''$, decreases by nearly an order of magnitude in shear rheology. Upon being compressed, with the bulk liquid being expelled, the average coordination number increases, and the corresponding clustering coefficient decreases, resulting in a more homogeneous network. Furthermore, nanoparticles can diminish the contact line pinning effect due to the liquid films, and the promoted liquid exchange leads to an averaged-out bridge volume~\cite{liu2024}.


Samples with nanoparticles inside the bridges lack the initial uniform shrinking stage in comparison to samples without, the fingering instability starts simultaneously
with the stretching. The competition between their instability and cavitation dynamics is similar to capillary suspensions in terms of time scale. Samples of similar sizes can be cavitation-dominated or fingering-dominated without a clear pattern, reflects the inherent variability in capillary suspension microstructure. Sample preparation steps, including mixing, loading, and slight compression during experiments can influence their fingering behaviors to varying degrees, making it hard to predict precisely. Nevertheless, we observed that larger initial diameters ($D_0$) consistently promote cavitation development, as demonstrated by three representative samples at different timestamps in supplementary Fig.~S3.

\begin{figure}[tbp]
    \centering
      \includegraphics[width=0.75\textwidth]{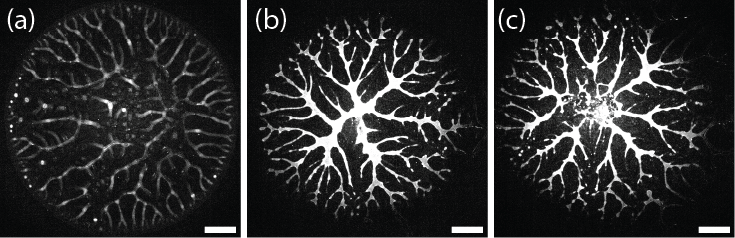}
      \caption{At $t = 18$~ms, patterns of (a) suspension (b) capillary suspension, and (c) capillary suspensions with nanoparticles with comparable initial diameters (32 $\pm$ 2~mm), dewetted with an acceleration of 50~m/s$^{2}$. The scale bars are 5~mm..}
      \label{fig:ThreeComparison}
    \end{figure}

The nanoparticle-caused differences might appear insignificant regarding instabilities, the focus lies however on the pattern morphology. Suspension, capillary suspension, and capillary nanosuspension with stable final patterns are shown in Fig.~\ref{fig:ThreeComparison}a to c, respectively. 
In Fig.~\ref{fig:ThreeComparison}a, the fingering branches are of similar thickness and dull with limited signal intensity. This indicated that most microparticles are detached from the plate during particle migration. With the attractive force provided by liquid bridges, the capillary suspension network resists retraction and the final pattern persists with high signal intensity, as shown in Fig.~\ref{fig:ThreeComparison}b. By adding nanoparticles, the outer branches become more discontinued, the inner branches are thinner and patchier with a higher tendency of cavitation (Fig.~\ref{fig:ThreeComparison}c). These morphological changes reflect an earlier breakage of nanoparticle-laden bridges, due to the combined effect of elevated restitution coefficient, reduced Hertzian contact, and eliminated contact line pinning.

\begin{figure}[tbp!]
    \centering
      \includegraphics[width=0.5\textwidth]{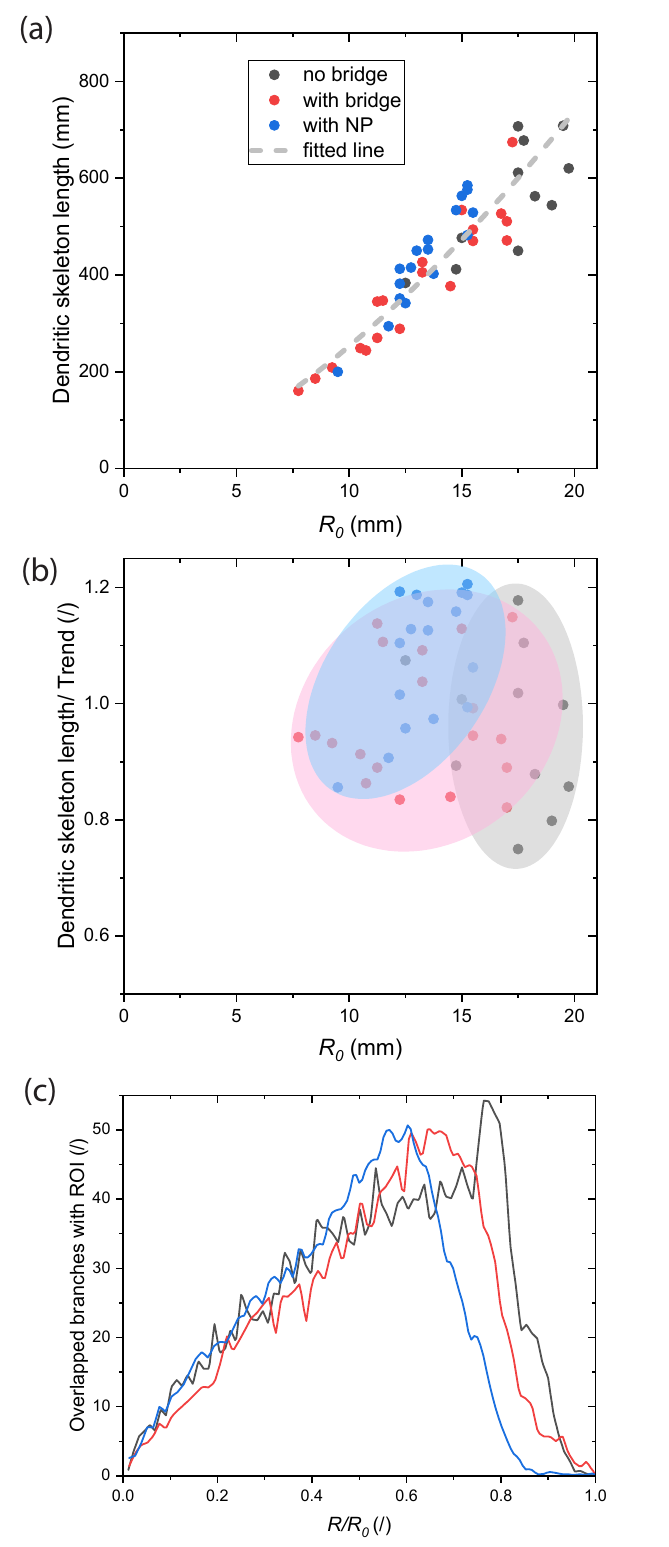}
      \caption{(a) Dendritic skeleton lengths are plotted against the initial sample radius for all three types of samples, the fitted power function is shown in a dashed gray line. (b) The ratio between the real dendritic skeleton length and the theoretical value. The shaded areas are to guide the eyes on data point groups. (c) Averaged intersecting branches with ROI for all three types of samples. The sample color corresponds to the legend shown in (a).}
      \label{fig:Overall_area_ratio_overlap}
    \end{figure}

Quantitatively, 0.01~vol\% nanoparticles have a more pronounced effect. The total dendritic length over radius and the ratios between the real values and theoretical fitted values are as shown in Fig~\ref{fig:Overall_area_ratio_overlap}a and ~\ref{fig:Overall_area_ratio_overlap}b. 
The power fitted line of the fingering length can be expressed as $L_{finger} = 143.6 R^{1.55} $, both units are in mm. The theoretical value has a power curve fitting of smaller than 2, indicating that in general the growth in dendritic length is smaller than the growth in sample area $4 \pi R^2$. Similar to the capillary suspension samples, the cavitation in capillary nanosuspensions also increases the fingering instability and this effect is more pronounced as the initial sample radius $R_0$ increases. Due to the preferred cavitation effect, the nanosuspension samples (blue) are mostly above the theoretical values. In Fig~\ref{fig:Overall_area_ratio_overlap}b, the total dendritic skeleton lengths are normalized over their initial radius, the higher the radius, the more deviated the distribution is from the fitted line. The capillary nanosuspensions are reflected by the points within the blue area, showing a positive linearity. This linearity also exists for normal capillary suspensions (red), though much less pronounced, indicated by the larger shadowed area. This narrow distribution of blue points fits perfectly with our previous observation. When nanoparticles are added to the capillary suspensions, they enhance the relative movement of the microparticles and promote liquid exchange, thus, resulting in a more homogeneous structure, particularly when subjected to external forces~\cite{liu2024}. Therefore, the homogeneous microstructure leads to a smaller sample-to-sample variation in the fingering pattern of capillary nanosuspension. On the other hand, the fingering ratios of the pure suspensions (gray) do not show any dependency regarding the initial sample radius, as it is more particle distribution related, the values differ from 0.7 to 1.2 randomly. 

In Fig.~\ref{fig:Overall_area_ratio_overlap}c, the averaged intersecting branch numbers are demonstrated over the normalized radius, despite the distinguished differences in fingering patterns, the maximal intersecting fingering numbers are all roughly 50. At the center region ($R/R_0 < 0.2$), capillary suspensions (red) have slightly lower intersecting branches than the pure suspension due to the large central bridge. The elastic network sometimes loses competition to the cavitation effect, and this is more pronounced with the presence of nanoparticles, causing the intersecting number to increase. Approaching the outer part near the menisci region at the three-phase contact line ($R/R_0 = 0.6-1$), the instability of the suspension (Fig~\ref{fig:noBridges_top}b and ~\ref{fig:noBridges_top}c) is strongly suppressed by the interparticle capillary force and the peak is shifted from $R/R_0 = 0.76$ to 0.64 With the addition of nanoparticles, the previously suppressed side branches are released approaching  $R/R_0 = 0.6$, the peak of capillary nanosuspensions (blue). The discontinued branches in capillary nanosuspension, caused by lubricated particle movement, can contribute to the strong decrease in intersecting fingering numbers at the outer part. 

\begin{figure}[tbp]
    \centering
      \includegraphics[width=0.6\textwidth]{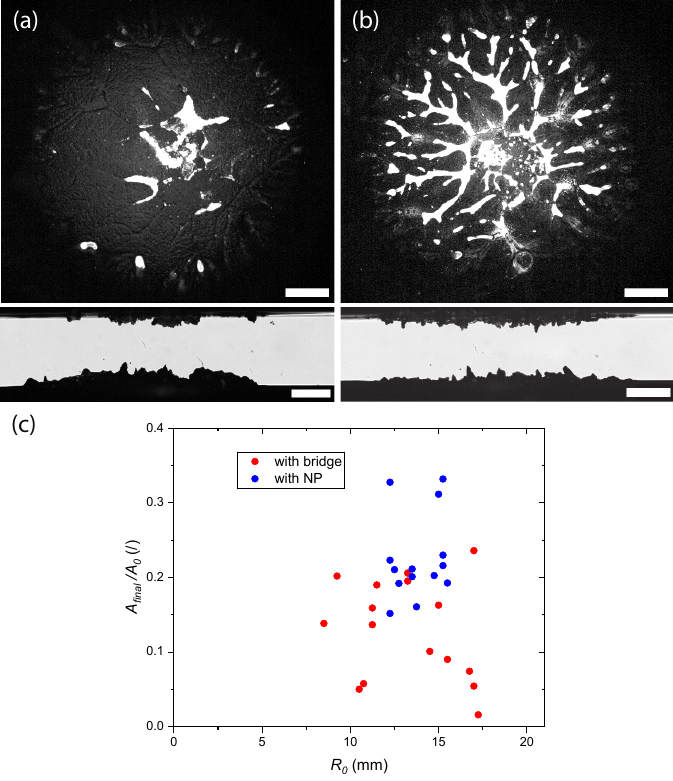}
      \caption{The residual of capillary suspensions (a) without (b) with nanoparticles from the top (up) and side (down) views, with an initial diameter $D_0$ of (a) 36~mm and (b) 33~mm, on the top glass. The samples are dewetted with an acceleration of 50~m/s$^{2}$, and the scale bars are 5~mm. (c) Ratios between residual area and initial area for capillary suspension with and without nanoparticles, from the top view.}
      \label{fig:oreology}
    \end{figure}

Nanoparticles contribute not only to the fingering pattern at the transient dewetting stage but also to the residual distribution. The residual distribution of capillary suspensions without (Fig~\ref{fig:1Bridges_top}b) and with (Fig~S3c) nanoparticles (both with cavitation pattern) is as shown in Fig~\ref{fig:oreology}a and ~\ref{fig:oreology}b respectively. 
A similar problem was raised by Owens~et~al.~\cite{owens2022}, when Oreo cookies were split into half, adhesive failure was dominant, and nearly 95~\% of the creme stayed on one wafer. This corresponds greatly to the situation of capillary suspension as shown in Fig~\ref{fig:oreology}a, where from the top view only 7.4~\% of the sample remained. The residual is mainly focused in the sample center where cavitation occurred, from the side view, an excessive amount of the sample stayed on the bottom plate as the microparticles are internally connected, and the attractive network pulls the gel structure from the top plate during dewetting. Nanoparticles on the other hand strongly promote the detachment of microparticles from each other, resulting in a more localized deposition. Based on the eased breakup of gel strings, adhesive failure can be turned into cohesive failure with the incorporation of nanoparticles, where the samples are more evenly distributed on both plates. The summarized residual area percentages are as plotted in Fig.~\ref{fig:oreology}c, from which it can be concluded that the normal capillary suspensions with no nanoparticles have a very pronounced adhesion which results in poor attachment of samples. Even for points with high residual area ratios of 20.3~\% and 23.6~\% on the top plate, the sample failure distribution is not even, a clear preferred deposition can be seen in Supplemental Fig.~S4a and S4b respectively. When the acceleration is tripled to 150~g/m$^2$, this cavitation effect is much stronger, which favors the deposition as shown previously (Supplemental Fig.~S5).



\section{Conclusions}

In this study, we utilized a modified lifted Hele-Shaw cell device with a fast-moving lower substrate, and we combined the imaging systems from the top and the side views to explore dewetting-induced fingering instability patterns and the splitting of the capillary suspensions with and without nanoparticles. 

Microparticles in pure liquid caused increased fingers and dendritic length due to enhanced particle interaction and these inclusions serve as nucleation sites for bubbles at low pressure. By adding secondary fluid, the bridges are formed and the total fingering number decreases, with a pronounced suppression at the center and the outer region, due to the introduction of a strong sample-spanning network with high elasticity. However, due to strong network heterogeneity, their fingering instability and cavitation effect are hard to predict, the fingering length fluctuates and the final residual is uneven. With the addition of only 0.01~vol\% nanoparticles, interesting phenomena have been observed. The capillary nanosuspensions' overall fingering instability becomes less sample-dependent with a high linear correlation between its initial radius. This aligns with our previous study, where nanoparticles lubricate the intermicroparticle movement and homogenize the sample distribution, indicated by lower clustering coefficients in comparison to the normal capillary suspensions~\cite{liu2024}. Meanwhile, the post-stretching residual of the gel structure can be strongly standardized by adding nanoparticles inside the liquid bridges. The elevated breakup between microparticle bridges can profoundly improve the sample deposition and change the failure type from adhesive failure, where the majority of the samples are stuck on one (random) plate, to cohesive failure, where the distribution is much more even.

The previous lifted Hele-Shaw cell studies were conducted on Newtonian fluids at quasi-equilibrium state~\cite{dias2013,kanhurkar2022,kanhurkar2019} with a few focusing on the high acceleration stretching~\cite{brulin2020,brulin2020b,moffatt2021}. The high-speed stretching case is relevant for industrial applications such as additive manufacturing, 3D printing, microfluidics, etc, where the dewetting procedure is essential.

 \FloatBarrier
 
\section*{Supporting Information}

Supplementary data on the gap height increase over time, the fingering instability evolution shown through normalized area and shrinkage rate, representative dendritic skeleton length with intersecting branches, and high-speed imaging of pattern formation at different accelerations and initial sample sizes.

\section*{Author Contributions}

\textbf{Lingyue Liu}: Conceptualization, Methodology, Investigation, Software, Writing – original draft. \textbf{Mete Abbot}: Conceptualization, Methodology, Investigation, Software, Writing – original draft.  \textbf{Philipp Brockmann}: Software, investigation, Writing – review \& editing. \textbf{Ilia V. Roisman}: Funding acquisition, Conceptualization, Methodology, Writing – review \& editing, Supervision. \textbf{Jeanette Hussong}: Funding acquisition, Conceptualization, Methodology, Writing – review \& editing, Supervision. \textbf{Erin Koos}: Funding acquisition, Conceptualization, Methodology, Writing – review \& editing, Supervision.

\section*{Acknowledgments}

This project has received funding from the European Union's Horizon 2020 research and innovation program under the Marie Skłodowska-Curie grant agreement No 955612 (NanoPaInt) and the International Fine Particle Research Institute (IFPRI). This research was supported by the German Scientific Foundation (Deutsche Forschungsgemeinschaft) in the framework of the SFB-1194 Collaborative Research Center “Interaction between Transport and Wetting Processes” (No 265191195), project A03.


\newpage

    \begin{figure}[tbp]
    \centering
      \includegraphics[width=\textwidth]{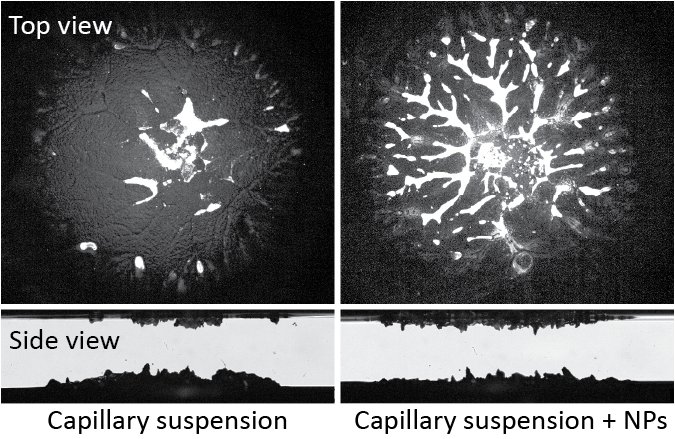}
      \caption{TOC Graphic}
      \label{fig:GraphicAbstract2}
    \end{figure}

\end{document}